\documentclass[twocolumn,aps]{revtex4}

\usepackage{dcolumn}
\usepackage{graphicx}

\begin{document}

\preprint{HEP/123-qed}
\title{All-electron $GW$ calculation based on the LAPW method:\\
application to wurtzite ZnO}
\author{Manabu Usuda and Noriaki Hamada\\}
\affiliation{
Department of Physics, Faculty of Science and Technology,
Tokyo University of Science,\\
2641 Yamazaki, Noda, Chiba 278-8510, Japan
}
\author{Takao Kotani}
\affiliation{
Department of Physics, Osaka University,
1-16 Machikaneyama, Toyonaka 560, Japan
}
\author{Mark van Schilfgaarde}
\affiliation{
Sandia National Laboratory, Livermore, CA U.S.A.
}
\date{\today}

\begin{abstract}
We present a new, all-electron implementation of the $GW$
approximation and apply it to wurtzite ZnO.  Eigenfunctions
computed in the local-density approximation (LDA) by the
full-potential linearized augmented-plane-wave (LAPW) or the
linearized muffin-tin-orbital (LMTO) method
supply the input for generating the Green function $G$ and 
the screened Coulomb interaction $W$.  A mixed basis is
used for the expansion of $W$,
consisting of plane waves in the interstitial region
and augmented-wavefunction products in the augmentation-sphere regions.  
The frequency-dependence of the dielectric function is computed
within the random-phase approximation (RPA), 
without a plasmon-pole approximation.  The Zn
$3d$ orbitals are treated as valence states within the LDA; both
core and valence states are included in the self-energy
calculation.  The calculated bandgap is smaller than experiment
by $\sim$1~eV, in contrast to previously reported $GW$ results.
Self-energy corrections are orbital-dependent, and push down
the deep O $2s$ and Zn $3d$ levels by $\sim$1~eV relative to the
LDA.  The $d$ level shifts closer to experiment but the size of
shift is underestimated, suggesting that the RPA overscreens
localized states.
\end{abstract}

\maketitle

\section{\label{sec:level1}Introduction}

Density-functional theory (DFT)
provides a foundation for modern electronic-structure calculations, and
the local-density approximation (LDA) is an efficient way to calculate
the ground-state properties of material.
However, the LDA eigenvalues should not necessarily identified with
the quasiparticle (QP) energies, although eigenvalue differences
are often used to describe the excited state.
Time-dependent DFT can in principle describe the excited state, but
a good approximation for the time-dependent
exchange-correlation kernel is not known.
The $GW$ approximation (GWA) of Hedin~\cite{Hedin} provides
a practical method to calculate the QP energy.
Hybertsen and Louie presented the first $GW$ calculation
for real materials in 1986~\cite{HL86}.
They employed eigenfunctions given by the LDA as input, using
additionally a pseudopotential approximation.

Several methods have since been developed within various
band-structure-calculation schemes~\cite{Arya00}.  Calculated
QP energies typically agree well with experiment, for many kinds
of materials.  However, various kinds of approximations in
addition to the $GW$ approximation itself are usually employed,
whose adequacy has not been well tested.  
Here we present a method that makes minimal approximations 
in addition to the $GW$ and random-phase approximations.
We start with the LDA eigenfunctions generated by the
full-potential linearized augmented-plane-wave (LAPW) \cite{FLAPW}
or a variant of the full-potential linearized muffin-tin-orbital (LMTO)
method~\cite{NFP,SSC}, where in either case eigenfunctions are
expanded in atomic-like local functions in the muffin-tin (MT)
sphere regions, and in plane waves in the interstitial region.
In order to treat the localized electrons accurately, the Coulomb
interaction $v$ and the screened Coulomb interaction $W$ are
expanded with a newly-developed mixed basis that consists of two
kinds of basis functions.
One is the product-basis function developed by Aryasetiawan and
Gunnarsson~\cite{AG94}, which is constructed from the products of
the local functions in the MT-sphere regions.
The other is the interstitial plane wave (IPW) that takes zero in
the MT-sphere regions and equals to the usual plane wave in the
interstitial region.  With the mixed basis, we can treat
localized electrons, even core electrons, on the same footing as
the other extended electrons.  The mixed basis is by construction
essentially a complete basis for the expansion of $W$; therefore,
for given eigenfunctions as input, our method can
produce reasonably well converged QP energies rather more
efficiently than a method that expands $W$ in plane waves alone.

The LMTO method employs smooth Hankel functions as envelope
functions~\cite{NFP}, which are smoother and more accurate
than the ordinary Hankel functions customary to the LMTO method.
About 100 envelope functions were employed in this calculation.
To check convergence, empty sphere was added to some of the
calculations, mainly to enlarge the basis set.  However, little
difference was found whether the empty sphere was added or not.

We apply this approach to wurtzite-type ZnO, whose valence bands
consist of extended O $2p$ and Zn $4s$ orbitals, and rather
localized Zn $3d$ and O $2s$ orbitals.
ZnO is important for optical device technology since the material
is optically transparent and can be doped with both electrons and
holes \cite{JTK99}.
Compared with most other II-VI and III-V compounds such as ZnS,
GaN, etc., the position of the cation $d$ levels are rather high,
and relatively close to the anion $p$-derived valence band maximum (VBM).  
The effect of the Zn $3d$ state is not negligible for
the various properties.  For example, the $3d$ state couples to
the VBM and pushes it upward, reducing the bandgap.  

This paper is organized as follows. Section \ref{sec. GW}
briefly describes the method.
In Sec. \ref{sec. results}, we show LDA and $GW$ results for ZnO,
and compare them to experimental data and previously reported $GW$
results.  A summary is given in Sec. \ref{sec. sum}.

\section{Overview of the $GW$ calculation}
\label{sec. GW}

The brief procedure of our $GW$ calculation is presented here;
it will be described in more detail elsewhere~\cite{future}.
In the Green's-function approach, the QP energy
$E_{{\bf k}n}$ and wavefunction $f_{{\bf k}n}({\bf r})$
of a many-electron system are given as the solution of the equation
\begin{eqnarray}
  \lefteqn{
    [E_{{\bf k}n}-T-V_{\rm H}({\bf r}) ]
    f_{{\bf k}n}({\bf r}) \qquad\qquad
  }\hspace{0.1cm}\nonumber \\
  & & {} \qquad
  - \int \Sigma({\bf r},{\bf r}',E_{{\bf k}n})f_{{\bf k}n}({\bf r}')d^3r'
  = 0,
  \label{eq. QP}
\end{eqnarray}
where $T$ is the kinetic-energy operator,
$V_{\rm H}$ is the Hartree potential plus the electrostatic potential
from nuclei, and $\Sigma$ is the self-energy.
In the $GW$ approximation,
the self-energy is written as
\begin{equation}
  \Sigma({\bf r},{\bf r}',\omega) =
  i\int_{-\infty}^{\infty} \frac{d\omega'}{2\pi}e^{i\omega'\delta}
  G({\bf r},{\bf r}',\omega+\omega')W({\bf r},{\bf r}',\omega').
  \label{eq. GW}
\end{equation}
It is convenient to divide $\Sigma$ into
$\Sigma = \Sigma_{\rm x} + \Sigma_{\rm c}$, with
$\Sigma_{\rm x}=iGv$ the bare exchange term, and
$\Sigma_{\rm c}=iGW_{\rm c}$
the correlation term.  Here $W_{\rm c}\equiv W-v$.

We take a perturbative approach to find $E_{{\bf k}n}$.
First, we solve the Kohn-Sham (KS) equation,
\begin{equation}
  [\epsilon_{{\bf k}n}-T
     -V_{\rm H}({\bf r})
     -V_{\rm xc}^{\rm LDA}({\bf r}) ]\psi_{{\bf k}n}({\bf r}) = 0,
  \label{eq. KS-LDA}
\end{equation}
where $V_{\rm xc}^{\rm LDA}({\bf r})$ is the exchange-correlation
potential in the LDA.
We assume that $\Sigma-V_{\rm xc}^{\rm LDA}$ is small
enough to be treated a first-order quantity.
Expanding $\Sigma(E_{{\bf k}n})$
around $\epsilon_{{\bf k}n}$,
we obtain to first order
\begin{equation}
  E_{{\bf k}n}
  \simeq
  \epsilon_{{\bf k}n}
  + Z_{{\bf k}n} [ \langle
    \psi_{{\bf k}n}|\Sigma(\epsilon_{{\bf k}n})
                      |\psi_{{\bf k}n}\rangle
  - \langle \psi_{{\bf k}n}|V_{\rm xc}^{\rm LDA}
    |\psi_{{\bf k}n}\rangle ],
  \label{eq. QPenergy2}
\end{equation}
where $Z_{{\bf k}n}$ is the renormalization factor defined by
\begin{equation}
  Z_{{\bf k}n} =
  [ 1- \langle \psi_{{\bf k}n}|(\partial \Sigma/\partial \omega)
    |_{\omega=\epsilon_{{\bf k}n}} |\psi_{{\bf k}n}\rangle ]^{-1} .
\label{eq. Z}
\end{equation}
We also estimate the first-order energy in the
Hartree-Fock approximation (HFA) through
\begin{equation}
  E_{{\bf k}n}^{\rm HFA} =
  \epsilon_{{\bf k}n} +
  \langle \psi_{{\bf k}n}|\Sigma_{\rm x}|\psi_{{\bf k}n} \rangle
- \langle \psi_{{\bf k}n}|V_{\rm xc}^{\rm LDA}
  |\psi_{{\bf k}n}\rangle ,
\label{eq. HFA}
\end{equation}
although the KS wavefunction may be different
from the HFA one, and, therefore, $E_{{\bf k}n}^{\rm HFA}$
is not the true self-consistent HFA value.

In the augmented-waves methods, space is divided into
the MT-sphere regions and the interstitial region.  In both LAPW
and the present LMTO method, the KS wavefunction is expanded as
\begin{equation}
  \psi_{{\bf k}n}({\bf r}) =
  \sum_{lm}\sum_{\beta=1,2}A_{alm\beta}({\bf r})
                \alpha_{alm\beta}^{{\bf k}n}  +
  \sum_{\bf G}P_{\bf G}^{\bf k}({\bf r}) z_n^{{\bf k}+{\bf G}}
  \label{eq. wf}
\end{equation}
where the atomic-like local function $A_{alm\beta}({\bf r})$ is defined by
\begin{equation}
 A_{alm\beta}({\bf r}) =
 \cases{
   \displaystyle \phi_{al\beta}(r)Y_{lm}(\hat{\bf r}) &
                 in the $a$-atom MT-sphere \cr
                 \cr
   \displaystyle 0 & otherwise \cr
   }
\end{equation}
with orthogonal radial wavefunctions $\phi_{al\beta}(r)$ ~($\beta=1$ or $2$),
and spherical harmonics
$Y_{lm}(\hat{\bf r})$.
The IPW $P_{\bf G}^{\bf k}({\bf r})$ is defined by
\begin{equation}
P^{\bf k}_{\bf G}({\bf r}) =
 \cases{
   \displaystyle 0 &
                 in the MT-sphere regions   \cr
                 \cr
   \displaystyle e^{i({\bf k+G})\cdot{\bf r}} &
                 in the interstitial region . \cr
   }
\end{equation}

The interactions,  $v$ and $W$, are well expressed
by the product of two KS eigenfunctions in our perturbative treatment.
The product
$\psi_{{\bf k}_1n_1}({\bf r})\psi_{{\bf k}_2n_2}({\bf r})$
is expanded by the product of two local functions,
$A_{al_1m_1\beta_1}({\bf r})A_{al_2m_2\beta_2}({\bf r})$,
in the MT-sphere regions,
and by the product of two plane waves,
$P_{{\bf G}_1}^{{\bf k}_1}({\bf r})P_{{\bf G}_2}^{{\bf k}_2}({\bf r})$,
in the interstitial region.
Following Aryasetiawan and Gunnarsson \cite{AG94} the complete
set of product functions is reduced by eliminating nearly
linearly dependent ones.
Taking the Bloch sum of the product functions,
we finally obtain the product-basis function expressed as
$B_{a\mu}^{\bf k}({\bf r})$ in the $a$-atom MT-sphere region.
The product of IPW's is also IPW in the interstitial region:
$P_{{\bf G}_1}^{{\bf k}_1}({\bf r})P_{{\bf G}_2}^{{\bf k}_2}({\bf r})
=P_{{\bf G}_1+{\bf G}_2}^{{\bf k}_1+{\bf k}_2}({\bf r})$.
Thus, we obtain a mixed basis
$\{ M_i^{\bf k}({\bf r}) \} \equiv
 \{ B_{a\mu}^{\bf k}({\bf r}), P_{\bf G}^{\bf k}({\bf r}) \}$
which is suitable for expansions of $v$ and $W$.
The index $i$ specifies a member of the basis and
runs through ${\bf G}$ and $a\mu$.

Because of the nonorthogonality of IPW's,
the overlap integral of the mixed-basis functions,
\[ S_{ij}\equiv \langle M_i^{\bf k}|M_j^{\bf k}\rangle ~, \]
is nonvanishing for $i\ne j$. We therefore define the dual-basis function:
\begin{equation}
 \tilde{M}_i^{\bf k}({\bf r}) \equiv
 \sum_{i'}M_{i'}^{\bf k}({\bf r})S_{i'i}^{-1}   ~~.
\end{equation}
The Coulomb interaction $v$ is expanded as
\begin{eqnarray}
  v({\bf r},{\bf r}') &=& \sum_{\bf k}^{\rm BZ}\sum_{ij}
       \tilde{M}_i^{\bf k}({\bf r})v_{ij}({\bf k})
       (\tilde{M}_j^{\bf k}({\bf r}'))^*        \nonumber \\
  v_{ij}({\bf k}) &=& \langle M_i^{\bf k}|v|M_j^{\bf k} \rangle ~~.
\label{eq. v}
\end{eqnarray}
The Coulomb matrix $v_{ij}({\bf k})$ can be calculated by
using the structure constants \cite{future}.

The self-energy is calculated by using
$\epsilon_{{\bf k}n}$,
$v_{ij}({\bf k})$ and
$\langle \psi_{{\bf q}n}|\psi_{{\bf q}-{\bf k}n'}\tilde{M}_i^{\bf k} \rangle$.
The diagonal part of $\Sigma_{\rm x}$ is given by
\begin{eqnarray}
\lefteqn{
 \langle \psi_{{\bf q}n}|\Sigma_{\rm x}|\psi_{{\bf q}n}\rangle
 }\hspace{0.1cm}\nonumber \\
 &=& - \sum_{\bf k}^{\rm BZ}\sum_{ij}\sum_{n'}^{\rm occ.}
   \langle \psi_{{\bf q}n}|\psi_{{\bf q}-{\bf k}n'}\tilde{M}_i^{\bf k}\rangle
   v_{ij}({\bf k})
   \langle \tilde{M}_j^{\bf k}\psi_{{\bf q}-{\bf k}n'}|\psi_{{\bf q}n}\rangle.
   \nonumber \\
\label{eq. SEX}
\end{eqnarray}
The non-interacting polarization function $D$
is expanded in the same manner as Eq.(\ref{eq. v}):
\begin{eqnarray}
\lefteqn{
 D_{ij}({\bf q},\omega) }\hspace{0.1cm}\nonumber \\
  &=&
  \sum_{\bf k}^{\rm BZ}\sum_n^{\rm occ.}\sum_{n'}^{\rm unocc.}
  \langle M_i^{\bf q}\psi_{{\bf k}n}|\psi_{{\bf k}+{\bf q}n'}\rangle
  \langle \psi_{{\bf k}+{\bf q}n'}|\psi_{{\bf k}n}M_j^{\bf q}\rangle
  \nonumber \\
  & & \times
  \Biggr[ \frac{1}
          {\omega-\epsilon_{{\bf k}+{\bf q}n'}+\epsilon_{{\bf k}n}+i\delta}
       -  \frac{1}
          {\omega+\epsilon_{{\bf k}+{\bf q}n'}-\epsilon_{{\bf k}n}-i\delta}
  \Biggr].
  \nonumber \\
  \label{eq. D}
\end{eqnarray}
We use the tetrahedron method for the Brillouin zone (BZ) summation in
Eq.(\ref{eq. D}) following Ref.\cite{RF75}.
The screened Coulomb interaction is given by
$W=(1-vD)^{-1}v$ in the random-phase approximation (RPA).
$W$ is also represented by the mixed basis.

The correlation part of the self-energy is calculated as
\begin{eqnarray}
\lefteqn{
 \langle \psi_{{\bf q}n}|\Sigma_{\rm c}(\omega)|\psi_{{\bf q}n}\rangle
 }\hspace{0.1cm}\nonumber \\
 &=&
 \sum_{\bf k}^{\rm BZ}\sum_{n'}^{\rm All}\sum_{ij}
 \langle \psi_{{\bf q}n}|\psi_{{\bf q}-{\bf k}n'}\tilde{M}_i^{\bf k}\rangle
 \langle \tilde{M}_j^{\bf k}\psi_{{\bf q}-{\bf k}n'}|\psi_{{\bf q}n}\rangle
 \nonumber \\
 & & \ \ \times \frac{i}{2\pi}\int_{-\infty}^{\infty}
         \frac{W_{ij}^{\rm c}({\bf k},\omega')}
         {\omega+\omega'-\epsilon_{{\bf q}-{\bf k}n'}\pm i\delta}d\omega',
 \label{eq. SEC}
\end{eqnarray}
where $-i\delta$ is taken for the occupied states and $+i\delta$ for the
unoccupied states.
We perform the frequency integration in Eq.(\ref{eq. SEC})
with a method devised by Aryasetiawan~\cite{Arya00}.

The evaluation of the exchange self-energy must be carried out carefully
since the Coulomb interaction $v_{ij}({\bf k})$ shows a singular behavior
$v_{ij}({\bf k}) \propto \frac{U_i^{0*}({\bf k})U_j^0({\bf k})}{|{\bf k}|^2}$
as ${\bf k} \to 0$, where $U_j^0({\bf k})$ denotes
the corresponding normalized eigenfunction.
The singularity also exists in $W$.
To handle the singular behavior,
we use the offset $\Gamma$-point method \cite{future}, which
is essentially equivalent to the method to integrate the
divergent part analytically~\cite{HL86}.

The QP energies are calculated with including
the core contributions through the following equation,
\begin{eqnarray}
  E_{{\bf k}n} &=& \epsilon_{{\bf k}n}
  + Z_{{\bf k}n}[\ \langle \psi_{{\bf k}n}|
    \Sigma_{\rm x }^{\rm core1} +
    \Sigma_{\rm xc}^{\rm core2+valence}
    |\psi_{{\bf k}n}\rangle \nonumber \\
  & & {} \qquad\quad
  - \langle \psi_{{\bf k}n}|V_{\rm xc}^{\rm LDA}(n_{\rm total})
    |\psi_{{\bf k}n}\rangle \ ],
\end{eqnarray}
where we divide the core states into two groups:
core1 is the deep core state, which affects the QP energies only through
$\Sigma_{\rm x}$, and core2 is the relatively shallow core,
which is treated on the same footing as the valence electrons.

\section{Results for Zinc Oxide}
\label{sec. results}

We first describe
the LDA calculation performed using the full-potential LAPW method.
The local exchange-correlation functional of Vosko, Wilk and Nusair \cite{VWN} is employed.
The space group of wurtzite ZnO is $P6_3mc$.
The lattice constant and the MT-sphere radii
are given in Table~\ref{tab. lattice}.
The angular-momentum in the spherical-wave expansion
is truncated at $l_{\mathrm{max}}=6$ and $l_{\mathrm{max}}=4$
for the potential and the wavefunction, respectively.
This $l_{\rm max}=4$ value is rather small, which gives errors
compared with more accurate calculations with $l_{\rm max}=7$, but
the differences of them are within 0.02 eV for the LDA band gap.
The energy cut-off of the IPW is 16Ry for the wavefunction.
We take 1152 $k$-points in the first Brillouin zone (BZ).
The Zn $(3d)^{10}(4s)^2$ and O $(2s)^2(2p)^4$ electrons are
treated as valence electrons.

The $GW$ calculation is performed with 32 $k$-points in the BZ.
The energy cut-off of the IPW is 10~Ry for the Coulomb matrix.
We treat 18 occupied bands and take into account 100 unoccupied bands.
When producing the product basis, we ignore the product functions including
$\phi_{al\beta=2}$ because the terms make small contributions.
The Zn $3p$ states, which is relatively shallow in the core states,
are chosen to be core2 electrons, i.e.,
the Zn $3p$ states are treated on the same footing as valence states
and taken into account for the calculation of the correlation part
of the self-energy. All the core and valence electrons are included
into the calculation of the exchange part of the self-energy.
In Sec.~\ref{sec. conv}, we check the convergence of the QP energies
in $k$-points, plane waves, unoccupied states,
and product functions.


\subsection{LDA}
\begin{figure}[ht]
  \includegraphics[width=\linewidth]{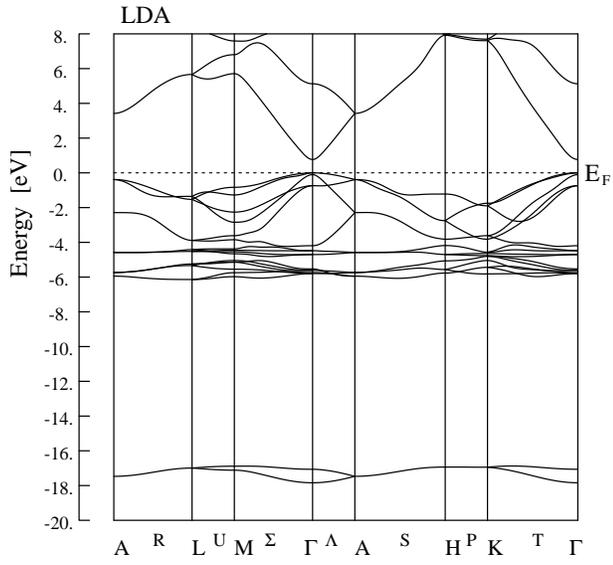}
  \caption{LDA band structure in wurtzite ZnO.}
  \label{LDA band1}
  \end{figure}

\begin{figure}[ht]
  \includegraphics[width=\linewidth]{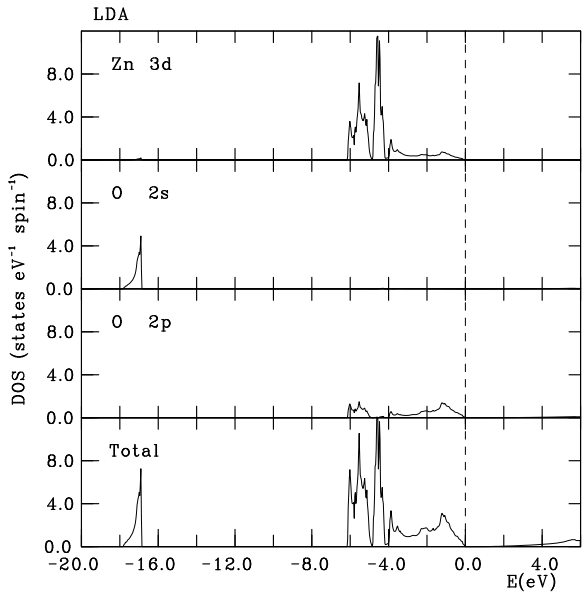}
  \caption{Density of states in wurtzite ZnO.}
  \label{LDA DOS1}
  \end{figure}

Unless otherwise stated, results in this section refer to the LAPW method.
The LDA band structure for ZnO is shown in
Fig.~\ref{LDA band1} and the density of states (DOS) is shown
in Fig.~\ref{LDA DOS1}. All energies are given with respect to the
top of the valence band.
Around $-$17 eV, we obtain two bands originating from the O $2s$ states.
The narrow bands between $-$6 and $-$4 eV consist mainly of the
Zn $3d$ orbitals, and the moderately dispersive bands from $-$4 
to 0 eV consist mainly of the O $2p$ orbitals.
Fig.~\ref{LDA DOS1} shows significant $p$-$d$ hybridization.
The Zn $3d$-derived bands are split into two groups, leading to
a double-peak structure in the DOS.
The lower peak is characterized by the strong $p$-$d$ hybridization.
The sharp upper peak between $-$4.8 and $-$4.2 eV has strong Zn $3d$ 
character and the hybridization with the O $2p$ states is very small.
The band gap opens between the 18th and 19th bands and
the fundamental gap of 0.77~eV is located at the $\Gamma$ point.
The lowest two conduction bands consist mainly of the Zn $4s$
orbitals. The energy levels at the $\Gamma$ point computed by 
the LAPW method are shown in Table~\ref{tab. QP}.
Also shown are the levels computed by the LMTO method.  
Agreement is excellent; the bandgaps are
nearly identical and the bands agree to within
0.1~eV over the entire valence bands.  A further check was
made using an entirely different kind of LMTO method
\cite{OFP}, and similar agreement was found.


The band dispersion is similar to
the result of the norm-conserving pseudopotential method
by Schr\"oer, Kr\"uger and Pollmann~\cite{SKP93}.
One important discrepancy is the magnitude of the band gap:
their fundamental gap is only 0.23 eV.
Other reported LDA gaps are 0.93 eV calculated with the
full-potential LAPW method \cite{MRPB95} and 1.15 eV within the
LMTO-ASA method \cite{OA00}.

\subsection{GWA}
\label{subsec. GWA}
\begin{figure}[ht]
  \includegraphics[width=\linewidth]{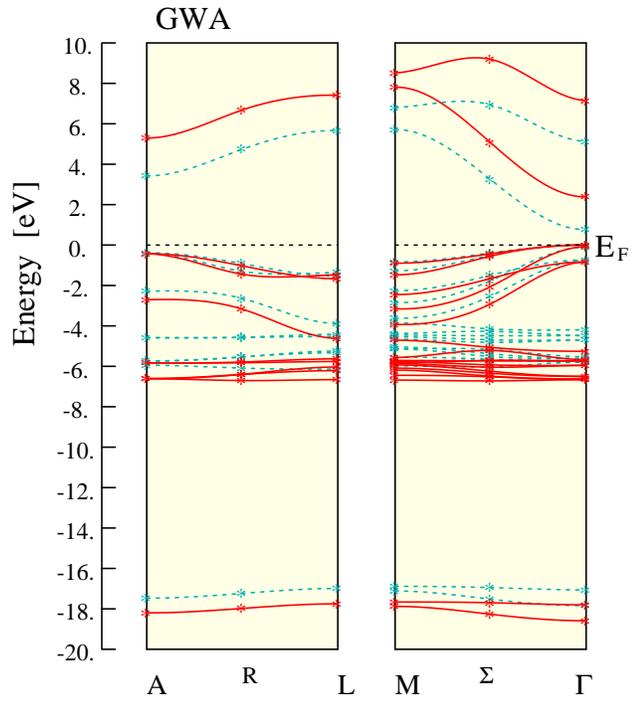}
  \caption{The $GW$ band structure (solid lines) and
  the LDA band structure (dashed lines) for wurtzite ZnO.
  The self-energy corrections are calculated at the points marked '$\ast$'.}
  \label{QP band}
  \end{figure}

The $GW$ band structure computed with LAPW input is shown in
Fig.~\ref{QP band} together with the LDA band structure.  In the
figure, all energies are given with respect to the top of the
valence band (the valence-band top in the GWA was shifted up
0.49~eV relative to the LDA).
We also show the LDA, $GW$ and HFA energies at the $\Gamma$ point
in Table~\ref{tab. QP}.
The first and second bands are the O $2s$ bands,
the 3-12th are mostly Zn $3d$ character,
bands 13-18 are of mostly O $2p$ character,
and bands 19-20 the Zn $4s$ conduction bands.
Because the Zn $3d$ bands and the O $2p$ bands overlap in energy,
the characterization of the band is more complicated:
e.g., at the $\Gamma$ point, the LDA 8th state  has an O $2p$ character,
while the 13th state has a Zn $3d$ character.

The self-energy correction is sensitive to
the character of the band, as seen from Table~\ref{tab. QP}.
The lowest O $2s$ bands are shifted by $\sim -1$ eV;
the Zn $3d$ bands by $\sim -1$ eV;
the conduction Zn $4s$ bands by $\sim +2$ eV.
The Zn $3d$ bandwidth shrinks,
while the O $2p$ bandwidth and the Zn $4s$ bandwidth are enhanced.
The O $2s$ bandwidth does not change.

The LAPW gaps are 0.77 eV, 2.44 eV and 11.39 eV in the LDA, GWA and HFA,
respectively, as Table~\ref{tab. QP} shows, for the 32 $k$-point mesh
employed.  The LMTO results are very similar.
The experimental gap is 3.44 eV~\cite{OA00}.
The large HFA gap reflects the neglect of screening the exchange
while the LDA gap is too small largely because it neglects
the nonlocality in the (screened) exchange altogether \cite{Maksimov}.
The $GW$ gap is closer to the experiment but still
smaller by $\sim$1 eV than the experimental gap.
Previously reported $GW$ gaps are somewhat larger.  Using a model $GW$
approach, Massidda et al. found a gap of 4.23 eV \cite{MRPB95} and
Oshikiri et al. found a gap of 4.28 eV \cite{OA00} using an all-electron
approach within the atomic spheres approximation.
The major part of the error in our gap energy is thought to be derived from
the overestimate of the dielectric function, originating from
underestimated LDA gap used to generate it.
We have found previously~\cite{SSC} that $GW$ gaps are systematically
underestimated in semiconductors, with the error increasing with ionicity.
Using an all-electron $GW$ implementation within the PAW method (albeit
including valence-only electrons in the $GW$ calculation), Arnaud and
Alouani \cite{AA00} have also noted a tendency for the $GW$ band gaps to be
smaller than the experimental ones.
For example, for Si, they find a gap of 1.00 eV.
Using a fully converged 512 $k$-point mesh,
our LAPW-$GW$ gap is 0.88~eV and our LMTO-$GW$ gap is 0.89~eV.
Wei Ku~\cite{WeiKu} has recently reported the LAPW-$GW$ gap of
0.85~eV using a 512 $k$-point mesh.



The centers of the O $2s$ band and the Zn $3d$ band are
$-$20.7 eV \cite{LPMKS74} and $-$8.81 eV \cite{Zn3d}, respectively, 
in experiment.
Table~\ref{tab. QP} shows that those $GW$ bands are
located higher by about 2 eV than the experimental positions.
The HFA result is rather closer to the experiment.
In our calculation, the dielectric function is overestimated
for the reasons listed below.
(1) In the RPA, the electric attractive force is neglected between
excited electron and hole; there is no restoring force for the polarization,
leading to the overestimation of the polarization function.
(2) The LDA band structure has a much smaller band gap than the experimental one.
The electron-hole pair excitation energy is smaller than the real system.
Since we use the LDA band structure to calculate the polarization function,
we overestimate it.
(3) The Zn $3d$ and O $2p$ bands are well
separated in experiment, while they partly overlap in the LDA.
The hybridization between the Zn $3d$ orbital and the O $2p$ orbital is
overestimated. The hybridization between the Zn $4s$ orbital and 
the O $2p$ orbital is also overestimated because of the smaller bandgap 
in the LDA. This property of the LDA wavefunction makes the charge transfer 
between Zn and O, leading to the overestimation of the polarization function.
All those procedures of our calculation causes the overestimation of the
dielectric function, and, therefore, the screened Coulomb interaction, $W$.
In order to improve our calculation in the latter two points of (2) and (3),
we need to perform a self-consistent calculation, that is, to use
the QP energies and wavefunctions to calculate
the screened Coulomb interaction $W$.

In Table~\ref{tab. GW}, we show the real part of $\Sigma$
at the $\Gamma$ point. In the table, we take notations,
$\Sigma_{{\bf k}n}^{\rm GWA}\equiv \langle \psi_{{\bf k}n}|
 \Sigma(\epsilon_{{\bf k}n})|\psi_{{\bf k}n} \rangle$ and
$\Sigma_{{\bf k}n}^{\rm LDA}\equiv\langle \psi_{{\bf k}n}|
 V_{\rm xc}^{\rm LDA}|\psi_{{\bf k}n} \rangle$;
the self-energy corrections are given by
$Z_{{\bf k}n}[\Sigma_{{\bf k}n}^{\rm GWA}-\Sigma_{{\bf k}n}^{\rm LDA}]$.
The normalization factor $Z_{{\bf k}n}$ is between 0.66 and 0.81,
which is comparable to simple metals or semiconductors.  As noted
before, the self-energy corrections are negative for the valence
bands and positive for the conduction bands and are larger for
the localized states.
Table~\ref{tab. Sigma}
shows the decomposition of $\Sigma_{{\bf k}n}^{\rm GWA}$
at the $\Gamma$ point to the core-exchange part
$\Sigma_{{\bf k}n}^{\rm xcore1}$,
the exchange part $\Sigma_{{\bf k}n}^{\rm x}$, and
the correlation part $\Sigma_{{\bf k}n}^{\rm c}$.
The contributions of the core2 electrons are included in
$\Sigma_{{\bf k}n}^{\rm x}$ and $\Sigma_{{\bf k}n}^{\rm c}$.
The exchange part $\Sigma_{{\bf k}n}^{\rm x}$ has a large discontinuity
across the Fermi level, leading to the wide gap in the HFA.
The correlation part $\Sigma_{{\bf k}n}^{\rm c}$ are positive
for the valence bands and negative for the conduction bands,
leading to the reduction of the bandgap from the HFA value.

\subsection{Convergence check}
\label{sec. conv}
The $GW$ band structures described in Sec.\ref{subsec. GWA}
have been calculated using 32 $k$-points in the BZ,
an IPW energy cutoff of 10~Ry for the Coulomb matrix,
100 unoccupied states, and
neglecting products including $\phi_{al\beta=2}$.
To check the convergence of the QP energies,
we have performed $GW$ calculations
with some different conditions.  The conditions we have used are
64 $k$-points 144 $k$-points, the energy cut-off of 16 Ry for the IPW, or
200 unoccupied states. The results are shown in
Table~\ref{tab. conv k}-\ref{tab. conv unocc.},
compared to the result in Sec.\ref{subsec. GWA}.
In any case, the errors are within 0.1 eV.
We have also performed the calculation with the products
including $\phi_{al\beta=2}$ for Zn.
The improvement is quite small, within 0.005 eV.
Our $GW$ result thus shows good convergence;
especially, a small number of IPW is enough to achieve a good result.

We have checked the influence of the Zn $3p$ core states.
In Sec.\ref{subsec. GWA}, we have treated the states as core2,
i.e. they treated on the same footing as valence states.
Now, we treat the Zn $3p$ states as core1 (and include the products
with $\phi_{al\beta=2}$ for Zn). The fundamental gap becomes 2.47 eV 
and the Zn $3d$ bands are shifted lower by $\sim$ 0.1~eV.
The effect of the Zn $3p$ electrons on the valence band through
the correlation term in the self-energy is about 0.1~eV.


\section{Summary}
\label{sec. sum}
We have presented a procedure for calculating the self-energy in the GWA
with the mixed-basis expansion based on the full-potential LAPW and LMTO
methods, and have applied the all-electron $GW$ calculation to wurtzite
ZnO.  The mixed-basis method works well for this system which has both
extended states and localized states; the $GW$ calculation has a good
convergence in various parameters, and can be performed on a
workstation-level computer.

The $GW$ bandgap of ZnO is smaller than experiment by $\sim$ 1 eV.
The self-energy correction is orbital dependent and
the localized O $2s$ and Zn $3d$ states are lowered by $\sim$ 1eV
relative to the LDA values, while still higher than experiment.
The $GW$ calculation overestimates the screening effect
for localized states such as the Zn $3d$ states,
because of the RPA and the LDA band structure.
These errors are apparently systematic in zincblende
semiconductors~\cite{SSC}.
We have indicated that a self-consistent calculation can
improve the result.

For a four-atom system having localized $3d$ orbitals such as
wurtzite ZnO, we can complete the $GW$ calculation with 32 
$k$-points in the BZ within three days by using a DEC alpha
21264 667MHz workstation.  Our $GW$ code is available in a web
site. The address is http://al1.phys.sci.osaka-u.ac.jp

\acknowledgments
We would like to thank Dr. Ferdi Aryasetiawan for many helpful discussions.
This work is supported by the Research Fellowships of the Japan Society
for the Promotion of Science for Young Scientists.
This work is also supported by a Grant-in-Aid for "Research for the Future"
Program from the Promotion of Science.
MvS was supported by the Office of Basic Energy
Sciences, under contract no. DE-AC04-94AL85000.




  \begin{table}[ht]
  \caption{ \label{tab. lattice}
     Crystal structure and MT radii for wurtzite ZnO.
     Atomic positions are $(\frac{1}{3}, \frac{2}{3}, 0)$ and
     $(\frac{2}{3}, \frac{1}{3}, \frac{1}{2})$ for Zn,
     and $(\frac{1}{3}, \frac{2}{3}, u)$ and
     $(\frac{2}{3}, \frac{1}{3}, u+\frac{1}{2})$ for O.
     We use experimental values of $a$ and $c$ given
     in Ref.~\cite{Lattice}, while the $u$-parameter of O is
     estimated within the LDA separately with the LAPW and LMTO calculations.
     The change in eigenvalues computed with the two different
     values of $u$ is small.  In some of the LMTO
     calculations, an empty sphere was added to enlarge the basis
     set.}
  \begin{ruledtabular}
  \begin{tabular}{cdd}
                      & \rlap{{\rm LAPW}}\quad  & \rlap{\rm LMTO}\quad      \\
    $a$ [\AA]         & 3.253\tablenotemark[1]  & 3.253\tablenotemark[1]    \\
    $c/a$             & 1.6025\tablenotemark[1] & 1.6025\tablenotemark[1]   \\
    $u$               & 0.3817                  & 0.3825                    \\ \hline
    Zn radius [a.u.]  & 2.0                     & 2.13                      \\
    O  radius [a.u.]  & 1.4                     & 1.70                      \\
    Es radius [a.u.]  &                         & 2.04                      \\
  \end{tabular}
  \end{ruledtabular}
  \footnotetext[1]{Reference \cite{Lattice}.}
  \end{table}

  \begin{table}[ht]
  \caption{\label{tab. QP}
  The LDA, $GW$ and HFA energies (in eV) at ${\bf k}=(0,0,0)$.
  All energies are given with respect to the top of the valence band.
  The LAPW-$GW$ and LMTO-$GW$ results use 32 and 216 $k$-points, 
  respectively. As shown in Tables~\ref{tab. conv k},
  \ref{tab. conv PW} and \ref{tab. conv unocc.}, the LAPW-$GW$ fundamental 
  gap would be 0.05 to 0.1~eV smaller if the combined effect of complete
  convergence in $k$-points, IPW and number of unoccupied states were taken
  into account.}
  \begin{ruledtabular}
  \begin{tabular}{lrrrrr}

            &\multicolumn{2}{c}{LDA}&\multicolumn{2}{c}{GWA}&  \multicolumn{1}{c}{HFA}  \\
   Band $n$  &  LAPW     & LMTO      &    LAPW   &    LMTO   &  LAPW     \\
  \tableline
      1      & $-$17.84  & $-$17.74  & $-$18.56  & $-$18.37  & $-$23.17  \\
      2      & $-$17.06  & $-$16.97  & $-$17.78  & $-$17.67  & $-$22.42  \\
             &           &           &           &           &           \\
     3,4     &  $-$5.80  &  $-$5.77  &  $-$6.62  &  $-$6.51  &  $-$9.95  \\
      5      &  $-$5.74  &  $-$5.70  &  $-$6.58  &  $-$6.49  &  $-$9.99  \\
     6,7     &  $-$5.62  &  $-$5.59  &  $-$6.46  &  $-$6.34  &  $-$9.80  \\
      8      &  $-$5.53  &  $-$5.61  &  $-$5.65  &  $-$5.59  &  $-$5.58  \\
     9,10    &  $-$4.70  &  $-$4.63  &  $-$5.90  &  $-$5.82  &  $-$9.87  \\
    11,12    &  $-$4.48  &  $-$4.42  &  $-$5.68  &  $-$5.61  &  $-$9.75  \\
             &           &           &           &           &           \\
     13      &  $-$4.19  &  $-$4.12  &  $-$5.20  &  $-$5.13  &  $-$8.92  \\
    14,15    &  $-$0.74  &  $-$0.79  &  $-$0.84  &  $-$0.87  &  $-$0.89  \\
     16      &  $-$0.10  &  $-$0.09  &  $-$0.08  &  $-$0.03  &  $-$0.00  \\
    17,18    &     0     &     0     &     0     &     0     &     0     \\
             &           &           &           &           &           \\
     19      &     0.77  &     0.78  &     2.44  &     2.44  &    11.39  \\
     20      &     5.13  &     5.12  &     7.19  &     7.25  &    17.13  \\

\end{tabular}
\end{ruledtabular}
\end{table}


  \begin{table}[ht]
  \caption{\label{tab. GW}
  $GW$ self-energies
  $\Sigma_{{\bf k}n}^{\rm GWA}$ at ${\bf k}=(0,0,0)$,
  together with the LDA exchange-correlation term
  $\Sigma_{{\bf k}n}^{\rm LDA}$ and
  the renormalization factor $Z_{{\bf k}n}$.
  The corrections are given by
  $Z_{{\bf k}n}[\Sigma_{{\bf k}n}^{\rm GWA} - \Sigma_{{\bf k}n}^{\rm LDA}].$}
  \begin{ruledtabular}
  \begin{tabular}{llllc}
   Band $n$ &
   $\Sigma_{{\bf k}n}^{\rm GWA}$ [eV] & $\Sigma_{{\bf k}n}^{\rm LDA}$ [eV] &
                           $Z_{{\bf k}n}$ & Corrections [eV] \\
  \tableline
    1   &   $-$22.09  &  $-$20.33  &  0.69  &  $-$1.22  \\
    2   &   $-$23.28  &  $-$21.45  &  0.66  &  $-$1.21  \\
        &             &            &        &           \\
   3,4  &   $-$35.25  &  $-$33.37  &  0.70  &  $-$1.32  \\
    5   &   $-$35.74  &  $-$33.83  &  0.70  &  $-$1.33  \\
   6,7  &   $-$35.74  &  $-$33.84  &  0.70  &  $-$1.33  \\
    8   &   $-$18.62  &  $-$17.74  &  0.70  &  $-$0.61  \\
   9,10 &   $-$41.70  &  $-$39.36  &  0.72  &  $-$1.69  \\
  11,12 &   $-$42.46  &  $-$40.11  &  0.72  &  $-$1.69  \\
        &             &            &        &           \\
    13  &   $-$40.66  &  $-$38.55  &  0.71  &  $-$1.50  \\
  14,15 &   $-$27.44  &  $-$26.65  &  0.75  &  $-$0.59  \\
    16  &   $-$27.93  &  $-$27.30  &  0.75  &  $-$0.47  \\
  17,18 &   $-$28.79  &  $-$28.13  &  0.75  &  $-$0.49  \\
        &             &            &        &           \\
    19  &   $-$12.87  &  $-$14.32  &  0.81  & \ \ 1.18  \\
    20  &   $-$12.45  &  $-$14.39  &  0.81  & \ \ 1.57  \\
  \end{tabular}
  \end{ruledtabular}
  \end{table}

  \begin{table}[ht]
  \caption{\label{tab. Sigma}
       Core-exchange part $\Sigma_{{\bf k}n}^{\rm xcore1}$,
       exchange part $\Sigma_{{\bf k}n}^{\rm x}$
       and correlation part $\Sigma_{{\bf k}n}^{\rm c}$ of
       the $GW$ self-energy at ${\bf k}=(0,0,0)$.
       The contributions of the core2 electrons are included in
       $\Sigma_{{\bf k}n}^{\rm x}$ and $\Sigma_{{\bf k}n}^{\rm c}$.}
  \begin{ruledtabular}
  \begin{tabular}{lccc}
  Band $n$ &
         $\Sigma_{{\bf k}n}^{\rm xcore1}$ [eV] &
         $\Sigma_{{\bf k}n}^{\rm x}$ [eV]      &
         $\Sigma_{{\bf k}n}^{\rm c}$ [eV]    \\
  \tableline
    1   &  $-$1.85  &  $-$29.84  & \ \ 9.60   \\
    2   &  $-$2.02  &  $-$30.81  & \ \ 9.55   \\
        &           &            &           \\
   3,4  &  $-$5.04  &  $-$38.51  & \ \ 8.29   \\
    5   &  $-$5.17  &  $-$38.93  & \ \ 8.36   \\
   6,7  &  $-$5.18  &  $-$38.87  & \ \ 8.31   \\
    8   &  $-$1.24  &  $-$22.57  & \ \ 5.20   \\
   9,10 &  $-$6.83  &  $-$43.73  & \ \ 8.85   \\
  11,12 &  $-$7.01  &  $-$44.39  & \ \ 8.95   \\
        &           &            &           \\
    13  &  $-$6.55  &  $-$42.75  & \ \ 8.65   \\
  14,15 &  $-$2.85  &  $-$29.97  & \ \ 5.39   \\
    16  &  $-$2.92  &  $-$30.31  & \ \ 5.30   \\
  17,18 &  $-$3.14  &  $-$31.02  & \ \ 5.37   \\
        &           &            &           \\
    19  &  $-$1.45  &  \ $-$8.27  & $-$3.15   \\
    20  &  $-$1.44  &  \ $-$6.97  & $-$4.04   \\
  \end{tabular}
  \end{ruledtabular}
  \end{table}

  \begin{table}[ht]
  \caption{\label{tab. conv k}
     Dependence on the QP energies on the number of $k$-points, 
     for representative states at the $\Gamma$ point.
     The calculation with 32 $k$ points
     is the same one in Table~\ref{tab. QP}.  The LMTO method
     was used to compute QP energies for finer meshes, up to 216
     $k$-points (6$\times$6$\times$6 mesh). The LMTO-$GW$ gap changed by
     $-$0.1~eV going from 64 to 216 $k$-points; assuming the same
     convergence in $k$-points within the LAPW method, we estimate 
     the converged LAPW-$GW$ fundamental gap to be 2.32~eV, keeping
     all other parameters fixed. The LAPW-$GW$ and LMTO-$GW$ energies 
     agree with each other to within 0.1~eV, with the error approximately 
     tracking differences in the LDA eigenvalues.}
  \begin{ruledtabular}
  \begin{tabular}{cccc}
  Band $n$ & 144 ${\bf k}$ & 64 ${\bf k}$ &  32 ${\bf k}$  \\
  \tableline
     1     &   $-$18.51  &  $-$18.54   &  $-$18.56   \\
     3     &   $-$6.60   &   $-$6.60   &   $-$6.62   \\
     8     &   $-$5.64   &   $-$5.62   &   $-$5.65   \\
    14     &   $-$0.84   &   $-$0.83   &   $-$0.84   \\
    18     &      0      &      0      &      0      \\
    19     &      2.35   &      2.42   &      2.44   \\
  \end{tabular}
  \end{ruledtabular}
  \end{table}


  \begin{table}[ht]
  \caption{\label{tab. conv PW}
     Dependence of the QP energies on the number of IPW used to make the
     the Coulomb matrix, for representative states at the $\Gamma$ point.
     Data with the 10~Ry cutoff is the same as in Table~\ref{tab. QP}.}
  \begin{ruledtabular}
  \begin{tabular}{ccc}
  Band $n$ & 16 Ry & 10 Ry  \\
  \tableline
     1     &  $-$18.55   &  $-$18.56   \\
     3     &   $-$6.59   &   $-$6.62   \\
     8     &   $-$5.64   &   $-$5.65   \\
    14     &   $-$0.84   &   $-$0.84   \\
    18     &      0      &      0      \\
    19     &      2.47   &      2.44   \\
  \end{tabular}
  \end{ruledtabular}
  \end{table}

  \begin{table}[ht]
  \caption{\label{tab. conv unocc.}
     Dependence of the QP energies on the number of
     the unoccupied states, for representative states at the $\Gamma$ point.
     The calculation with 100 unoccupied states is the same as
     in Table~\ref{tab. QP}.}
  \begin{ruledtabular}
  \begin{tabular}{ccc}
  Band $n$ & 200 unocc. & 100 unocc. \\
  \tableline
     1     &  $-$18.56   &  $-$18.56   \\
     3     &   $-$6.54   &   $-$6.62   \\
     8     &   $-$5.62   &   $-$5.65   \\
    14     &   $-$0.83   &   $-$0.84   \\
    18     &      0      &      0      \\
    19     &      2.45   &      2.44   \\
  \end{tabular}
  \end{ruledtabular}
  \end{table}


\end{document}